\documentclass{sf2a-conf2018}
\usepackage{graphicx}
\usepackage{longtable}  
\usepackage{hyperref}
\usepackage[]{natbib}  
\usepackage[T1]{fontenc}
\usepackage{epstopdf}

\def\BibTeX{{\rm B\kern-.05em{\sc i\kern-.025em b}\kern-.08em
    T\kern-.1667em\lower.7ex\hbox{E}\kern-.125emX}}
\bibpunct{(}{)}{;}{a}{}{,}  %%%%%%%%%%%%%  A&A bibliography style
\newcommand{\kms}{{\mathrm{km~s^{-1}}}}

\newcommand{\teff}{{\mathrm{Teff}}}
\newcommand{\logg}{{\mathrm{\log g}}}

\begin{document}

\TitreGlobal{SF2A 2018}

%%-----------------------------------------------------------------
%%      the top matter
%%

\title{Elemental abundances of HD 87240, member of the young open cluster NGC 3114}

\runningtitle{HD 87240}

\author{V.Wellenreiter}\address{LESIA, UMR 8109, Observatoire de Paris Meudon, Place J.Janssen, Meudon, France}

\author{R.Monier}\address{LESIA, UMR 8109, Observatoire de Paris Meudon, Place J.Janssen, Meudon, France}

\author{L.Fossati}\address{Space Research Institute, Austrian Academy of Sciences, Schmiedlstrasse 6, 8042, Graz, Austria}

%% IF Author3 has the same affiliation than Author1:
%\author{C.\,E. Author3$^1$}

%% IF Author3 has its own affiliation:
%\author{C.\,E. Author3}\address{Dept. of Chess, University of Games, 35101 Las Vegas, Monaco} 

%% Keep this line, even if the page will be settled afterwards.
\setcounter{page}{237}

%%-----------------------------------------------------------------

\maketitle

%%-----------------------------------------------------------------
%%        The abstract
%% 
%%  Warning!  within the abstract:
%%  - do not use macros. 
%%  - do not use commands like: \cite, \citet, \citep ... etc.

\begin{abstract}
We have determined the abundances of several chemical elements in the atmosphere of HD 87240, a member of the young open cluster NGC 3114.
We have used the code ATLAS9 to compute a model atmosphere for HD 87240 for the effective temperature and surface gravity derived from Stromgren's photometry
A grid of synthetic spectra has been computed using SYNSPEC49 and adjusted to the UVES spectrum of HD 87240 to determine the abundances of several chemical elements using the latest critically evaluated atomic data from NIST.
We find underabundances of light elements and large overabundances of heavy elements, in particular of the Rare Earths, platinum and mercury which suggest that this star is a Chemically Peculiar late B star of the SiPtHg type.
\end{abstract}

%% Insert the keywords (to appear in the ADS indexing)
%% Keywords must be separated by a comma

%%-----------------------------------------------------------------

\section{Introduction}
%%---------------------

HD 87240 is a member of the young open cluster NGC 3114 classified as an Ap Si star. The age of NGC 3114 is estimated to be 160 million years, the distance derived from Gaia DR2 parallax of HD 87240 is about 2208 parsecs and its color excess is small  (E(B-V)=0.007 mag).
\cite{Saffe09} published a first abundance analysis for 16 elements.
In this work we derive new abundances or upper limits for HD 87240 using updated and  recent atomic data. 

\section{Observations and reduction}

We did not observe HD 87240 ourselves but fetched the spectrum from the ESO archive.
HD 87240 has been observed on 26 October 2017 using the high resolution (R =75000) mode of UVES.
One 1100 minutes exposure was secured with a $\frac{S}{N}$ ratio of about 270 at 5000 \AA.

\section{Atmospheric model and synthesis spectra}

The effective temperature and surface gravity of HD 87240  were first evaluated using Napiwotzky et al's (1993) UVBYBETA calibration of Stromgren's photometry.
The found effective temperature $\teff$ is 13319 $\pm$ 250 K and the surface gravity $\logg$ is 3.75 $\pm$ 0.25 dex. 
A plane parallel model atmosphere assuming radiative equilibrium, hydrostatic equilibrium and local thermodynamical equilibrium has been first computed using the ATLAS9 code \citep{Kurucz92}, specifically the linux version using the new ODFs maintained by F. Castelli on her website\footnote{http://www.oact.inaf.it/castelli/}. The linelist was built starting from Kurucz's (1992) gfhyperall.dat file  \footnote{http://kurucz.harvard.edu/linelists/} which includes hyperfine splitting levels.
This first linelist was then upgraded using the NIST Atomic Spectra Database 
\footnote{http://physics.nist.gov/cgi-bin/AtData/linesform} and the VALD database operated at Uppsala University \citep{kupka2000}\footnote{http://vald.astro.uu.se/~vald/php/vald.php}.
A grid of synthetic spectra was then computed with a modified version of SYNSPEC49 \citep{Hubeny92,Hubeny95} to model the lines. The synthetic spectrum was then convolved with a gaussian instrumental profile and a parabolic rotation profile using the routine ROTIN3 provided along with SYNSPEC49.
We derived a projected apparent rotational velocity $v_{e} \sin i =  7.5 \kms$ by modelling the Mg II triplet at 4480 \AA .

\section{Results and comparison with previous results}

We have derived abundances for 39 elements, namely the difference of the absolute abundance $\log ({X \over H})$ with the corresponding solar abundance. A representative minimum error of $\pm$ 0.15 dex is adopted. We find that HD 87240 displays underabundances in the light elements He, Mg, Al, S. It has solar abundances for C, N, O , Sc, V. Silicon is overabundant by a factor of 10. HD 87240 has large overabundances (larger than 5 times solar) in several very heavy elements: Ce, Pr, Nd, Eu, Dy, Ho (by about $10^3$ times the solar abundances) and in Pt and Hg. The heaviest element Pt and Hg are the most overabundant (by about $10^4$ the solar abundances). The abundances are collected in Table 1. 
The results obtained mostly differ from the results of \citep{Saffe09}.\\ 
For a few elements, the abundances are the same (Carbon, Iron, Chronium...) but they are quite different for several others (Ytrium, Cerium, Europium...).
The improved quality of atomic data probably accounts for the different results. Spectral variability due to spots could also account for the different abundances. %%-------------------

%\begin{figure}[h!]
% \centering
% \includegraphics[width=0.5\textwidth]{Fig_WM.pdf}     %% Note the ABSENCE of the extension .pdf , .eps or .ps  !
%  \caption{The abundance pattern of HD 87240}
%  \label{fig1}
%\end{figure} 

%%
%% Example of single figure
%%

%%
%% Example of two figures side by side
%%

% Table of abundances here

%***** exemple de tableau  (supprimer une ou 2 colonnes si necessaire ****************

%---- Line identifications 

\begin{table}
\centering
\begin{tabular}{|l|l|l|l|p{3cm}|}
 \hline
  \multicolumn{4}{|c|}{\textbf{Determined abundances in HD 87240}}   \\ \hline
  \multicolumn{1}{|c|}{Element} &
  \multicolumn{1}{|c|}{Abundance} &
  \multicolumn{1}{|c|}{Solar abundance} &
   \multicolumn{1}{|c|}{Comment}        \\ \hline
He  & -1.77 & -1.07 & \\
C   & -3.44 & -3.48 &  \\
N   & -4.02 (u.l)  & -4.02 &   \\
O   & -3.13 & -3.17 &   \\
Na  & -5.63 & -5.67  &  \\
Mg  & - 4.94 & -4.42  &   \\
Al  & -6.53 & -5.53  &   \\
Si  & -4.25 & -4.45  &    \\
P   & -6.25 & -6.55  &    \\
S   & -5.19 & -4.67  &    \\
Ca  & -4.99 & -5.64  &    \\
Sc  & -8.83 & -8.83  &     \\
Ti & -5.49  & -6.98  &      \\
Cr & -4.95  & -6.33  &    \\
Mn & -5.91  & -6.61  &    \\
Fe & -3.43  & -4.50  &     \\
Co  & -7.08 (u.l) & -7.08 &   \\
Ni  & -5.75 (u.l) & -5.75 &   \\
Cu  & -7.79 (u.l) & -7.79 &    \\
Ga  & -7.48  & -7.12 &  \\
Sr  & -6.67  & -9.03 &   \\
Y   & -7.86  & -9.16 &    \\
Zr  & -7.29  & -9.40 &     \\
Ba  & -9.87  & -9.87 &     \\
\hline
La  & -10.83 (u.l) & -10.83 & \\
Ce  & -7.38   & -10.42  &  \\
Pr  & -8.39   & -11.29  &  \\
Nd  & -7.10   & -10.50  &   \\
Sm  & -10.99 (u.l) & -10.99 & \\
Eu  & -8.45   & -11.49  &   \\
Gd  & -10.88 (u.l) & -10.88 & \\
Tb  & -12.10 (u.l) & -12.10 & \\
Dy  & -8.23  & -10.86  &    \\
Ho  & -8.74  & -11.74  &    \\
Er  & -7.59  &  -11.07 &   \\
Yb  & -10.92 (u.l) & -10.92 & \\
\hline
Pt  & -5.90  & -10.20 &     \\
Hg  & -6.87  & -10.91  &     \\  
\hline
\end{tabular}
\caption{Abundance analysis for HD 87240}  
\end{table}

%*********** transit Identifications ***************

\section{Conclusions}
%%--------------------
The derived abundance pattern of HD 87240 departs strongly from the solar composition which definitely shows that HD 87240 is not a superficially normal late B star but is definitely a new CP star. We have already reported on the discovery of 6 new CP stars of the HgMn type among late B-type stars in \cite{Monier15}, \cite{Monier2016} and \cite{Monier2018}. 
HD 87240 has overabundances of both the rare earths and of Pt and  Hg and its effective temperature and surface gravity place it among the Bp Si and the HgMn stars. Hence we propose that HD 87240 be reclassified as a Bp SiPtHg star (not as an Ap Si as it currently is).

% Optional acknowledgements
% -------------------------

%%-----------------------------
%%   Bibliography
%%-----------------------------
%%
%% The reference list should contain all the references cited in the text, ordered alphabetically by surname (with
%% initials following). If there are several references to the same first author, they should be entered according
%% to the following scheme:
%% 1. One author: chronologically
%% 2. Author, one co-author: alphabetically by co-author, then chronologically
%% 3. Author, two or more co-authors: chronologically.
%%
%% Please note that for papers that have more than five authors, only the first three should be given, followed
%% by "et al."
%%
%% The format for references is the one adopted by A&A (see the example below).
%%
%% To set the reference list in the proper A&A format, we encourage you to use BibTEX and the natbib
%% package instead of the standard 'thebibliography' environment.
%%

%% The following lines are required when using BibTEX (strongly encouraged!):
%\nocite{*}
%\bibliographystyle{aa.bst}  % A&A bibliography style file (aa.bst)
%\bibliography{sf2a-template.bib} % your references in file: Yourfile.bib

%
\end{document}